# The Measurement Gap in the Automation of EU Law: Benchmarking Doctrinal Legal Reasoning under the EU AI Act

**Michèle Finck**[1]

**Abstract**

Large language models now produce legal text of at least median quality, yet no existing benchmark can evaluate whether they perform doctrinal legal reasoning, which forms the interpretive core of legal work, rather than the ancillary, paralegal tasks that most current "legal AI" evaluations measure. [2] This measurement gap is not only methodological but legal: the EU AI Act makes "appropriate accuracy" a binding requirement for high-risk AI used in the judicial domain, yet that requirement cannot acquire operational content without the very doctrinal-reasoning benchmark the field lacks.

More than three years have passed since ChatGPT, built on GPT-3.5, thrust large language models into public consciousness. Even in legal circles, the acronym "LLM" now evokes neural networks more readily than the *Legum Magister* ("LL.M.") degree with which it was once exclusively synonymous. These models have already permeated legal research, practice, and pedagogy. Any academic who has marked take-home written work knows that they now routinely produce output of median quality; law firms have begun to capture the efficiency gains of generative AI on routine tasks; and a debate has emerged over how best to integrate these tools into judicial processes. Recent advances have spurred experimentation across every industry, including law. These are clear signs of a period of rapid evolution at the intersection of technology and law.

This paper proceeds from the premise that these developments will soon provoke a hard question: whether AI, and LLMs in particular, can perform the very core of legal work, namely doctrinal legal reasoning. The paper does not empirically settle that claim. Rather, it argues that continued model advancement may soon force the question, and that the field currently lacks the tools to answer it. Indeed, existing benchmarks test predominantly for tasks ancillary to legal work, specifically paralegal tasks or performance on law exams, rather than for doctrinal reasoning itself. It further contends that Article 15 of the AI Act turns this measurement gap into a legal one, since the "appropriate" accuracy required of high-risk judicial AI cannot be operationalised without doctrinal-reasoning benchmarks, making their development a constructive obligation on the Commission under Article 15(2). Beyond this diagnosis, the paper makes a second, constructive contribution. Written from the standpoint of a lawyer rather than a computer scientist, it offers not a concrete benchmark but a taxonomy of the failure modes that any benchmark genuinely testing doctrinal reasoning in EU law would need to avoid.

---

[1] Chair of Law and Artificial Intelligence and Director, CZS Institute for Artificial Intelligence and Law, University of Tübingen.

[2] This terminology is used with caution here as to the lawyer, the expression connotes the lawfulness of AI rather than its use in the legal domain.

These questions matter. Should models eventually cross the doctrinal reasoning threshold, the ramifications could be tectonic. The impact would extend far beyond the efficiency of law firms, the integrity of university examinations, or the human production of legal research. It would reach the very foundations of our democratic structures, which are built around, and governed by, the *human* production, interpretation, and enforcement of legal norms. It is therefore time to begin thinking about how to measure whether AI can do this work. Benchmarks are not unproblematic, and they should not be the only tool brought to bear in this debate. Indeed, the difficulty of matching doctrinal legal reasoning to benchmarks forces the question of other, more suitable, measurement instruments. Notwithstanding, it is high time we started thinking about them, given that they are the instrument machine learning has traditionally used to measure its own progress.

## I. Introduction

I vividly recall one of my early meetings with one of my doctoral supervisors. As a first-year doctoral candidate, I was enrolled in a mandatory legal research and methodology course. Having recently completed a research master's degree where I wrote a thesis without giving methodology much thought, I found myself confused by the variety of approaches presented to us. At our next meeting, I asked my supervisor whether I needed to incorporate elements that would have been novel to me, such as historical analysis, interviews or empirical data. She replied with assurance, saying something along the lines of: "Don't worry! You just keep doing doctrinal legal research. It is perfectly suited to your topic."[3] I left the meeting reassured. I took away the implicit message that my intuitive approach, namely reading various sources and weaving them into a narrative of how they did or did not fit together to then ponder related implications, was sufficient. Admittedly, however, I did not *really* stop to question what "doctrinal research" entailed. It was only when I taught an LLB class introducing legal systems and methods that the issue resurfaced. Even then, my approach remained largely intuitive. Recently, however, I have come to think that we may soon require a more explicit, more articulated, and — perhaps most uncomfortably for any lawyer — a more *quantified* engagement with the question of what doctrinal legal reasoning really is. This is so as we are now confronted with algorithms capable of producing legal text of at least median quality, and we have no agreed criteria with which to evaluate whether what they produce constitutes doctrinal reasoning or merely a surface-level approximation thereof.

This paper makes two diagnostic claim and one constructive contribution. The first diagnostic claim is that we cannot presently tell whether a model reasons doctrinally or merely reproduces the surface of legal language, because the benchmarks through which machine learning measures its progress test paralegal tasks such retrieval, classification, extraction, exam-style question answering or performance on law school exams rather than doctrinal reasoning itself. The second diagnostic claim is that the AI Act, in fact, requires the creation of doctrinal reasoning benchmarks. The constructive contribution, offered from the standpoint of a lawyer rather than a computer scientist, is a taxonomy of the failure modes that any benchmark genuinely testing doctrinal reasoning in EU law would need to detect.

The argument proceeds as follows. Section II identifies four structural features of doctrinal legal reasoning, namely internalism, normativity, contestability and coherence. Section III shows that each is intensified in EU law. Section IV argues that doctrinal reasoning benchmarks are now necessary on two grounds. First, epistemic, to settle a capability dispute that intuition can no longer resolve, and second, legal, because the EU AI Act's accuracy requirement for high-risk judicial AI cannot acquire operational content without them, making their development a constructive obligation on the Commission under Article 15(2). Section V surveys existing benchmarks for AI in the legal domain and shows that none tests doctrinal reasoning in EU law. Section VI makes a first contribution towards the creation of such benchmarks in setting out a failure taxonomy. Section VII concludes.

[3] This, by no means, does not imply that she has not deeply engaged with other methodological approaches. See, by way of example, Elizabeth Fisher, 'Imagining Method in Administrative Law Scholarship' in Carol Harlow (ed), *A Research Agenda for Administrative Law* (Edward Elgar 2023).

## II. Doctrinal Legal Reasoning

Given its focus on LLMs, this paper concerns itself with doctrinal legal *reasoning* rather than the broader practice of doctrinal legal *research*. While tightly connected, the two are distinct. Doctrinal legal research is the practice of producing systematised, justified accounts of what the law requires. Doctrinal legal reasoning is the core cognitive activity driving that practice: the process by which a lawyer, judge, or academic determines what a legal norm requires by synthesising relevant sources, articulating a defensible interpretation, and integrating it into the wider legal system, often against a fact pattern or hypothetical.

While the methodological literature is largely framed around doctrinal research (as this is what scholars most explicitly address), the four structural features this section identifies are ultimately features of the reasoning process rather than the published artifact. They serve as the criteria the legal community uses to judge whether a doctrinal argument has been competently made.

Doctrinal legal research is the oldest and most widely used methodological approach in law.[4] It has been labelled "the definitive form of legal scholarship".[5] Unlike other methodological approaches to law, such as socio-legal studies, critical legal studies, or law and economics, doctrinal research is used not only by academics but also by judges and practitioners. Doctrinal legal research is also prescriptive as it feeds back into the interpretation of legal sources and the drafting of legislation, shaping the law it purports to describe.[6] It follows that doctrinalism forms a crucial tool for anyone who has reason to construct an account of what the law requires. Its centrality to the law as a discipline and to the legal professions explains why an AI system capable of doing it well would carry consequences far exceeding those of, say, a system specialised in case law synthesis.

There is, however, no definition of doctrinal reasoning that commands consensus. This definitional challenge partly rests on the fact that approaches vary across jurisdictions. For instance, common law traditions tend toward historical approaches while civil law systems are more inclined towards conceptual and philosophical perspectives.[7] Doctrinal legal analysis furthermore assumes different nuances depending on the area of law in question.[8] This lack of definitional precision is enabled by the fact that academics and practitioners alike internalise the method largely through apprenticeship rather than through methodological instruction.[9]

[4] This is so as, as has been argued above and will be seen further below, doctrinal legal analysis not only also is used by judges and lawyers but because it also often forms a necessary starting point for other methodological approaches in law.
[5] W.T. Murphy and S. Roberts, `Introduction' (1987) 50 Modern Law Review 677, 667, quoted in Douglas Vick "Interdisciplinarity and the Discipline of Law" (2004) 31 Journal of Law and Society 163, 177.
[6] Stefan Vogenauer, 'An Empire of Light? II: Learning and Lawmaking in Germany Today' (2006) 26 Oxford Journal of Legal Studies 627.
[7] Mathias Siems and Daithí Mac Síthigh, 'Mapping Legal Research' (2012) 71 Cambridge Law Journal 651.

Richard Posner has described its object as "the messy work product of the judges and legislators", which "requires a good deal of tidying up, of synthesis, analysis, restatement, and critique".[10] These are tasks that are "intellectually demanding", requiring "not only brains and knowledge and judgment, but also Sitzfleisch", and that remain "of inestimable importance to the legal system and of greater social value than much esoteric interdisciplinary legal scholarship".[11] This paper does not advance a precise definition of doctrinal legal reasoning. Rather, it surveys the methodological literature to distill four structural features that any benchmark purporting to measure doctrinal competence must capture, namely (i) internalism; (ii) normativity; (iii) contestability and (iv) coherence.

First, doctrinal legal reasoning is *internal*. It adopts the perspective of a participant in the legal order rather than that of an external observer.[12] This can be contrasted with social-scientific approaches that ask "questions about the law in practice, of legal institutions at work in society rather than legal rules existing in a social, economic, and political" context.[13] Internalism is not a denial that external context matters and many doctrinal scholars in fact move between internal and external perspectives.[14] Internalism is, rather, a methodological commitment to take the law's own argumentative structures and standards of correctness as the starting point of inquiry. This implies that (i) legal sources are the exclusive avenue for altering rules and principles of law; (ii) legal sources are intelligible, coherent and consistent; and (iii) the success of any doctrinal account is contingent on legal sources alone.[15] The starting question for doctrinal reasoning is, accordingly "what is the law?".[16]

Second, doctrinal legal reasoning is *normative* because "legal doctrine is interpretative, that legal interpretation is inherently normative and that therefore legal doctrine is normative".[17] The normative nature of law presupposes an approach that is different from descriptive or empirical approaches.[18] The doctrinal scholar speaks from inside the system, using its normative vocabulary as binding. According to Hart, the law's normativity rests on a fundamental accepted norm, which he calls the rule of recognition.[19] Indeed, doctrinal reasoning can describe positive law as a system of valid norms, and argue about what it requires, while presupposing the legal point of view without personally endorsing it.[20] The law inherently has a "claim to correctness".[21] On the strict-internalist view, normative evaluation is something performed alongside doctrinal reasoning rather than within it. Pursuant to this perspective, doctrinal accounts "evaluate the attractiveness of articulated rules and principles exclusively on the basis of legal sources" and are in that sense "normatively inert", in that "the answers they furnish are not impacted by the desirability (moral or otherwise) of the rules and principles posited by legal sources".[22] A different strand of the literature, however, considers doctrinal standards to be evaluative rather than descriptive. Pursuant to this account, doctrinalism takes a position on what the law requires and is judged by the quality of the reasons that position rests on. Jan Smits has argued, accordingly, that doctrinal research often serves three distinguishable but interrelated aims: description (what does the law say?), prescription (what should it say?), and justification (why this rule rather than that?).[23] It follows that even within an internalist methodology, doctrinal reasoning involves evaluative judgment about which of several candidate readings of contested material best fits the legal system. That normative judgment may be implicit or explicit, conscious or unconscious, but it is always present in some form.

[8] Regarding administrative law, see further Elizabeth Fisher, 'Imagining Method in Administrative Law Scholarship' in Carol Harlow (ed), *A Research Agenda for Administrative Law* (Edward Elgar 2023) .
[9] Terry Hutchinson and Nigel Duncan, "Defining and Describing What We Do: Doctrinal Legal Research" (2012) 17 Deakin Law Review 83, 99 ("the doctrinal method is often so implicit and so tacit that many
working within the legal paradigm consider that it is unnecessary to verbalise the process".)
[10] Richard Posner, 'In Memoriam: Bernard D Meltzer (1914–2007)' (2007) 74 University of Chicago Law Review 435–38, 437.
[11] Ibid.
[12] Christopher McCrudden, 'Legal Research and the Social Sciences' (2006) 122 Law Quarterly Review 632, 648.
[13] Ibid.
[14] See further Martijn W Hesselink, 'A European Legal Method? On European Private Law and Scientific Method' (2009) 15 European Law Journal 20.
[15] Stefan Theil, 'Carefully Tailored: Doctrinal Methods and Empirical Contributions' (2025) 45 Oxford Journal of Legal Studies 1047, 1047.
[16] Terry Hutchinson, 'Vale Bunny Watson? Law Librarians, Law Libraries and Legal Research in the Post-Internet Era' (2014) 106 Law Library Journal 579, 584. See also Martha Minow, 'Archetypal Legal Scholarship: A Field Guide' (2013) 63 Journal of Legal Education 65, 65.
[17] Anne Ruth Mackor, 'Explanatory Non-Normative Legal Doctrine: Taking the Distinction between Theoretical and Practical Reason Seriously' in Mark van Hoecke (ed), *Methodologies of Legal Research* (Hart 2011) 45.

Third, doctrinal legal reasoning is *contestable*. A well-executed doctrinal argument produces not certainty but a strong defensible position that anticipates and engages with objections.[24] The underlying reason, as HLA Hart famously argued, is that legal rules have an "open texture", meaning that they admit of more than one defensible interpretation and this is not a defect of the system but a structural feature of it.[25] This is, in part, related to the fact that the law gives normative force to essentially contested concepts such as equality, dignity or fairness.[26] Even so, doctrinal reasoning constrains the set of admissible interpretations, which must rest on patterns of reasoning that the legal community recognises as valid. Indeed, legal systems consist of underlying rules and principles that bind judges even in so-called "hard cases", and that even contested questions admit of a defensible best reading rather than merely of multiple defensible ones.[27] Contestability, however, must be distinguished from indeterminacy. The doctrinal tradition proceeds from the more modest premise that legal questions, though genuinely arguable, remain rationally tractable.[28] Contestability is far more than a theoretical feature of law but is rather visible across the entire legal landscape. Research has, for instance, revealed immense variance in how law exams are graded.[29]

---

[18] Hans Kelsen, *Reine Rechtslehre* (Deuticke 1960). See further Carsten Heidemann, "Hans Kelsen's Normativism" (Cambridge University Press 2022).
[19] HLA Hart, *The Concept of Law* (Oxford University Press 1961) p. 91-107.
[20] See further Joseph Raz, *The Concept of a Legal System* (Clarendon Press 1979) 234-38.
[21] Robert Alexy, *Theorie der juristischen Argumentation* (Suhrkamp 1991).
[22] Theil (n16) at 1049. See further also Tarunabh Khaitan and Sandy Steel, 'Theorizing Areas of Law: A Taxonomy of Special Jurisprudence' (2023) 28 Legal Theory 325, 335.
[23] Jan M Smits, 'What is Legal Doctrine? On the Aims and Methods of Legal-Dogmatic Research' in Rob van Gestel, Hans-W Micklitz and Edward L Rubin (eds), *Rethinking Legal Scholarship: A Transatlantic Dialogue* (Cambridge University Press 2017) 207, 218–19. See also Rob van Gestel and Hans-Wolfgang Micklitz, 'Why Methods Matter in European Legal Scholarship' (2014) 20 European Law Journal 292, 311.
[24] McCrudden (n 3) 633. Indeed, even the internalism and normativity features outlined above have been contested. See further Anne Ruth Mackor's "Explanatory Non-Normative Legal Doctrine. Taking the Distinction between Theoretical and Practical Reason Seriously" in Mark van Hoecke "Methodologies of Legal Research" (Hart 2011) 45.
[25] HLA Hart, *The Concept of Law* (Oxford University Press 1961) ch 7.
[26] W B Gallie, 'Essentially Contested Concepts' (1956) 56 Proceedings of the Aristotelian Society 167.
[27] Ronald Dworkin, *Taking Rights Seriously* (Duckworth 1977); and Ronald Dworkin, 'Is There Really No Right Answer in Hard Cases?' in *A Matter of Principle* (Harvard University Press 1985) 119.

Finally, doctrinal legal reasoning is shaped by *coherence*. It does not aim merely at locally correct statements of discrete rules but rather to integrate those rules into a systematically ordered account.[30] Stated differently, norms need to "hang together" and make sense as a whole.[31] Consistency is hence not the mere absence of contradiction but rather the positive property of propositions hanging together so as to constitute an intelligible whole.[32] It is coherence that does the decisive work in hard cases. Where two readings are each consistent with the local rule, the doctrinal interpreter prefers the one that better fits the principles underlying the relevant body of law as a whole.

### III. Doctrinal Legal Reasoning in EU Law

Doctrinal reasoning takes on different nuances across jurisdictions and areas of law.[33] Focusing on EU law, this paper argues that these four features are both present and intensified.[34] Doctrinal legal reasoning in EU law is the activity of working out what a supranational legal norm requires: reading the Treaties, secondary legislation, CJEU case law and the other sources whose relevance varies by area and articulating a defensible interpretation on that basis. Its goal is "seeking answers to particular legal problems or to systematizing the state of the law".[35] It is, inter alia, performed in the legal services of the EU institutions, in Advocate General opinions, by practitioners, the CJEU and national courts applying EU law, as well as in academic writing. In EU law scholarship, there is an ongoing discussion regarding the methods used to study EU law.[36] Indeed, the methodological literature on EU legal interpretation is extensive but has not produced a definition of doctrinal reasoning that demands consensus.[37] This section does, accordingly, not provide a definition. It rather takes the four features identified above and shows how each is present, and indeed sharpened, in the supranational legal order.

[28] HLA Hart, *The Concept of Law* (Oxford University Press 2012) ch 7; Neil MacCormick, *Legal Reasoning and Legal Theory* (Oxford University Press 1978); and Robert Alexy, *A Theory of Legal Argumentation* (Ruth Adler and Neil MacCormick trs, Oxford University Press 1989).

[29] Clemens Hufeld, 'Jede Korrektur eine andere Note: Quantitative Untersuchung der Objektivität juristischer Klausurbewertungen' (2024) 11 ZDRW 59.

[30] Neil MacCormick, "Coherence in Legal Justification" in A. Peczenik et al (eds) "Theory of Legal Science" (Springer 1984).

[31] Ibid, 235.

[32] Robert Alexy and Aleksander Peczenik, 'The Concept of Coherence and Its Significance for Discursive Rationality' (1990) 3 Ratio Juris 130. See further also Joseph Raz, 'The Relevance of Coherence' (1992) 72 Boston University Law Review 273

[33] Mathias Siems and Daithí Mac Síthigh, 'Mapping Legal Research' (2012) 71 CLJ 651; Elizabeth Fisher, 'Imagining Method in Administrative Law Scholarship' in Carol Harlow (ed), *A Research Agenda for Administrative Law* (Edward Elgar 2023).

[34] See also van Gestel and Micklitz 294.

[35] Vincent Réviellère, 'Ongoing Controversies over Methods in EU Law' (Verfassungsblog 7 May 2025) https://verfassungsblog.de/ongoing-controversies-over-methods-in-eu-law/, last accessed 12 June 2026.

[36] Réviellère (n 36).

[37] See, by way of example, Koen Lenaerts and José A Gutiérrez-Fons, *Les méthodes d'interprétation de la Cour de justice de l'Union européenne* (Bruylant 2020); Gunnar Beck, *The Legal Reasoning of the Court of Justice of the EU* (Hart 2012); Gerard Conway, *The Limits of Legal Reasoning and the European Court of Justice* (CUP 2012); Giulio

Doctrinal reasoning, it was argued above, is *internal*. In EU law, internalism is not merely a feature of legal reasoning; it is constitutive of the legal order itself. The first reason is constitutional. The "inside" from which the EU doctrinal interpreter reasons did not pre-exist the case law; it was constituted by it. *Van Gend en Loos* declared a "new legal order" whose subjects include individuals, and *Costa* established that the validity of its norms cannot be assessed by reference to national law.[38] The criteria of validity, the hierarchy of sources and the boundary of the order are fixed from within.[39] Internalism in EU law hence requires an at least relative disregard for competing perspectives on EU law in national legal orders.[40] The second reason why internalism is constitutive of EU law is semantic internalism.[41] Concepts of EU law have autonomous meaning, which can be distinct from the meaning of the same concept in other jurisdictions.[42] This interpretation method is necessary to ensure the uniform application of EU law across all Member States and must account for the purpose and context of the concept in EU law.[43] For example, internalism mandates that legal concepts such as that of the "worker" or the "consumer" be interpreted solely in accordance with the meaning EU law ascribes to them.[44] AI for doctrinal legal reasoning in EU law must respect these features. Yet the training corpuses of LLMs blend concepts from different jurisdictions whose representation of a legal term is anchored in the statistical distribution of its uses in training data, which is overwhelmingly drawn from English-language, US-jurisdiction legal materials. [45]

Doctrinal reasoning in EU law is explicitly *normative*. The most distinctive interpretive feature of EU law is its teleological character, which can be more or less pronounced depending on the specific legal question that must be answered.[46] As Pierre Pescatore observed long ago, the Court has assigned priority to the teleological method because the Treaties' purpose-driven functionalism

---

Itzcovich, 'The Interpretation of Community Law by the European Court of Justice' (2009) 10 German Law Journal 537.

[38] Case 26/62 *Van Gend en Loos* ECLI:EU:C:1963:1; Case 6/64 *Costa v ENEL* ECLI:EU:C:1964:66.

[39] Opinion 2/13 ECLI:EU:C:2014:2454; Case C-284/16 *Achmea* ECLI:EU:C:2018:158.

[40] *Solange I*, BVerfGE 37, 271 (1974); *Solange II*, BVerfGE 73, 339 (1986); BVerfG, Judgment of 5 May 2020 (*PSPP*), BVerfGE 154, 17.

[41] While this is also true for some national legal orders, this is not necessarily the case. For instance, Luxembourg and Belgian courts look towards French interpretations of the same legal concepts due to the shared legal tradition. See further Konrad Zweigert & Hein Kötz, 'An Introduction to Comparative Law' (Oxford University Press 1998) Part I.

[42] Case 327/82 *Ekro* ECLI:EU:C:1984:11, para 11. Case 283/81 *CILFIT v Ministero della Sanità* ECLI:EU:C:1982:335, para 19.

[43] Ibid.

[44] On the concept of the "worker", see further Case 75/63 *Hoekstra* [1964] ECLI:EU:C:1964:19, para 1 and Case 66/85 *Lawrie-Blum* [1986] ECLI:EU:C:1986:284, paras 12-16. For the notion of the consumer, see Case C-464/01 Johann Gruber v Bay Wa AG [2005] ECLI:EU:C:2005:32, para 31.

[45] Jesse Dodge and others, 'Documenting Large Webtext Corpora: A Case Study on the Colossal Clean Crawled Corpus' (Proceedings of the 2021 Conference on Empirical Methods in Natural Language Processing, 2021) and Chris Wendler, Veniamin Veselovsky, Giovanni Monea and Robert West, 'Do Llamas Work in English? On the Latent Language of Multilingual Transformers' (Proceedings of the 62nd Annual Meeting of the Association for Computational Linguistics, 2024).

[46] Indeed, while teleological interpretation is crucial to understanding the development and current operation of EU law, the Court does not rely on it exclusively as it is often unnecessary, such as when addressing highly detailed legal questions within secondary legislation.

required it to give concrete expression to general and open-ended notions in light of the constitutional objectives the Union must pursue.[47] Joseph Weiler likewise drew attention to the "living political matrix" of EU law, that is to say "the interactions between norms and norm-making, constitution and institutions, principles and practice, and the Court of Justice and the political organs".[48] This purposive approach for instance explains why recitals are used as interpretational guides to substantive Treaty provisions.[49] The implication for doctrinal measurement is direct. A language model is an engine of descriptive distribution: prompted about a provision, it samples from the spread of readings weighted by their prevalence in its training data and shaped by its post-training, rather than reporting any single "correct" interpretation.[50] But the normativity of EU doctrinal reasoning means that the dominant reading is not, for that reason, the correct one. Rather, correctness tracks the better argument towards the order's objectives.

Doctrinal reasoning in EU law is also *contestable*. First, contestability in EU law takes on additional forms that exceed the general contestability of doctrinal legal reasoning that was outlined above such as linguistic contestability. EU law is multilingual, meaning that many sources exist in twenty-four equally authentic language versions, and one language version cannot be relied on as the sole interpretive basis or to override the others.[51] Where courts are made aware of divergences between language versions, they must take those divergences into account in their reasoning. Multilingualism hence multiplies defensible readings of EU law, increasing contestability. Second, EU law institutionalises contestability procedurally through Article 267 TFEU. The reference for preliminary rulings procedure explicitly acknowledges that even the highest national courts may disagree on different readings of EU law.[52] Reasoning doctrinally therefore requires reasoning contextually rather than treating case law as a flat and timeless corpus.

Finally, *coherence* matters greatly in EU law. One expression of this relates to the interpretation of legal norms. Indeed, in the supranational legal order, each provision "must be placed in its context and interpreted in the light of the provisions of [EU] law as a whole", regard being had to its objectives and state of evolution.[53] Coherence in EU law also relates to its hierarchy of norms.

---

[47] Pierre Pescatore, 'Les objectifs de la Communauté européenne comme principes d'interprétation dans la jurisprudence de la Cour de justice' in *Miscellanea W J Ganshof van der Meersch* vol 2 (Bruylant 1972) 325, 327–29.

[48] Joseph H H Weiler, 'The Transformation of Europe' (1991) 100 Yale Law Journal 2403, 2409.

[49] Maarten den Heijer, Teun van Os van den Abeelen and Antanina Maslyka, 'On the Use and Misuse of Recitals in European Union Law' (Amsterdam Law School Research Paper No 2019-31, 2019) https://ssrn.com/abstract=3445372 accessed 12 June 2026 and Tadas Klimas and Jūratė Vaičiukaitė, 'The Law of Recitals in European Community Legislation' (14 July 2008) Legal Studies Research Paper Series https://ssrn.com/abstract=1159604 accessed 12 June 2026.

[50] Christoph Engel and Richard McAdams, 'Asking GPT for the Ordinary Meaning of Statutory Terms' [2024] University of Illinois Journal of Law, Technology & Policy 235

[51] See Case 283/81 CILFIT v Ministero della Sanità ECLI:EU:C:1982:335, para 18 and C-561/19 Consorzio Italian Management e Catania Multiservizi and Catania Multiservizi ECLI:EU:C:2021:799, para 41-50.
For an interpretation, see further Imelda Maher, 'The CILFIT Criteria Clarified and Extended for National Courts of Last Resort Under Art 267 TFEU' (2022) 7 European Papers 265.

[52] C-561/19 Consorzio Italian Management e Catania Multiservizi and Catania Multiservizi ECLI:EU:C:2021:799, para 47 ("it is only where, with the help of the interpretation criteria set out in paragraphs 40 to 46 above, a national court or tribunal of last instance concludes that there is no circumstance capable of giving rise to any reasonable doubt as to the correct interpretation of EU law that that national court or tribunal may refrain from referring to the Court a question concerning the interpretation of EU law and take upon itself the responsibility for resolving it").

[53] *CILFIT* (n 8) para 20

Indeed, EU law is characterized by its layered constitutional structure. This comprises the Treaties but also the Charter of Fundamental Rights has the same legal value as the Treaties.[54] In addition, general principles of EU law, which are derived from the common constitutional traditions of the Member States and the ECHR operate as an uncodified constitutional source.[55] These instruments form substantive norms and EU secondary legislation must be read consistently with these sources.[56] This layered infrastructure is hard for LLMs to handle.[57] Indeed, the hierarchy is invisible at the level of text as nothing on the face of a directive's article discloses that its permissible readings are constrained from above, which is why a system that ingests the order as a flat corpus of documents is missing precisely the architecture that does the doctrinal work. Coherence is not only something EU doctrinal reasoning seeks but something the order is constitutionally committed to preserving. The unity and coherence of the legal order are values the Court treats as structural, bound up with its autonomy.[58]

The paper has so far set out what doctrinal legal reasoning consists in and why it forms the interpretive core of legal practice. The following section argues that recent advances in artificial intelligence now make it necessary to ask how that competence can be measured in EU law.

This article arises from my own experimentation with LLMs in late March 2026, directed at their capacity for doctrinal legal reasoning. My conclusion was that current models cannot, yet reason doctrinally. Others may reasonably disagree, and as models improve, such disagreement will grow harder to settle by impression. As the question is becoming more pressing evaluation instruments tailored specifically to doctrinal legal reasoning in EU law are needed.

## IV. From Intuition to Instrument: The Epistemic and Regulatory Necessity for Doctrinal-Reasoning Benchmarks

The instrument machine learning would prescribe is the benchmark.[59] A benchmark is a standardised combination of a dataset of inputs with target outputs and a scoring metric, conceptualised as representing a task, against which models are tested and ranked.[60] The paradigm, institutionalised as the "common task framework", has been credited with much of the field's progress[61], and it structures both academic publication and commercial development: a new model is presented to the community in the form of its scores on the relevant benchmarks. Benchmarks are thus how the ML community settles capability disputes that intuition cannot resolve and, for all their documented shortcomings, what would be reached for to assess a model promising doctrinal competence. A deeper analysis of the shortcoming of benchmarks falls outside the scope of this paper. [62] I examine benchmarks because, as will be seen below, the benchmark is the measurement instrument the AI Act singles out. The focus on benchmarks is also pragmatic: it rests on the assumption that the doctrinal-reasoning question will force itself upon us before any widely-accepted alternative methodological tool becomes available. That said, the discussion that follows can itself be read as supporting a more sceptical conclusion, namely that benchmarks are, in fact, not a suitable instrument for measuring a model's capacity for doctrinal legal reasoning at all due to the difficulty – and maybe impossibility – of agreeing on the constituent features of doctrinalism and measuring them. That, however, is an argument for another occasion.

[54] Article 6(1) TEU.

[55] Article 6(3) TEU.

[56] Takis Tridimas, *The General Principles of EU Law;* Case C-617/10 *Åkerberg Fransson* EU:C:2013:105

[57] Nelson Liu and others, 'Lost in the Middle: How Language Models Use Long Contexts' (2024) 12 Transactions of the Association for Computational Linguistics 157. Emily Bender and Alexander Koller, 'Climbing towards NLU: On Meaning, Form, and Understanding in the Age of Data' in Proceedings of the 58th Annual Meeting of the Association for Computational Linguistics (Association for Computational Linguistics 2020) 5185; Matthew Dahl and others, 'Large Legal Fictions: Profiling Legal Hallucinations in Large Language Models' (2024) 16 Journal of Legal Analysis 64; Brandon Waldon and others, 'Large Language Models for Legal Interpretation? Don't Take Their Word for It' (2025) 114 Georgetown Law Journal 115.

[58] Opinion 2/13 (n 6) paras 174–178.

[59] For an introduction to benchmarks, see Moritz Hardt, '1 – Introduction' in The Emerging Science of Machine Learning Benchmarks (2025) <https://mlbenchmarks.org/01-introduction.html> accessed 12 June 2026.

[60] Inioluwa Deborah Raji and others, 'AI and the Everything in the Whole Wide World Benchmark' (NeurIPS Datasets and Benchmarks 2021) arXiv:2111.15366; Percy Liang and others, 'Holistic Evaluation of Language Models' (2022) arXiv:2211.09110.

[61] David Donoho, '50 Years of Data Science' (2017) 26 Journal of Computational and Graphical Statistics 745.

This section outlines two overarching reasons why doctrinal-reasoning benchmarks for EU law are needed. The first is methodological: without them, the progress of models cannot be assessed (A). The second is legal: the AI Act, I argue, requires such benchmarks to be built for a specific class of high-risk AI systems (B).

### A. Tracking Model Advances in Doctrinal Legal Reasoning

There are various jurisdiction-independent reasons why it is necessary to measure LLM performance for doctrinal legal reasoning. First, at present, views whether LLMs can or cannot reason doctrinally rest on impression and personal experimentation.[63] That footing is unstable. Model capability has improved on a steep and, so far, sustained curve, tracking increases in scale and, more recently, in inference-time reasoning. [64] At the same time, the surface errors that make an LLM's failure to reason easy to spot, most visibly hallucination, have on many tasks receded over successive generations, even if this trend is not uniform.[65] As models improve, the disagreement between those who think the doctrinal legal reasoning threshold has been crossed and those who do not will harden, as the evidence on each side remains anecdotal and non-comparable.

Second, without measurement, competence cannot be told apart from its simulation. The most distinctive risk of these systems in the legal domain is not obvious error, which is easy to spot, but fluent surface that masks absent reasoning.[66] A model can produce a confident, well-formed legal answer whose underlying reasoning is missing or wrong, and existing ML products for the legal domain do so at material rates.[67] At present, there are no quantified measurement instruments able to distinguish a model whose representations track doctrinal architecture from one that reproduces its surface. Until that measurement exists, every strong claim about these systems, such as the vendor's that they reason like lawyers and the sceptic's that they cannot, rests on assertion rather than evidence.

[62] For an overview of these shortcomings, see further Maria Eriksson and others, 'Can We Trust AI Benchmarks? An Interdisciplinary Review of Current Issues in AI Evaluation' (2025) arXiv:2502.06559; Inioluwa Deborah Raji and others, 'AI and the Everything in the Whole Wide World Benchmark' (35th Conference on Neural Information Processing Systems (NeurIPS) Track on Datasets and Benchmarks, 2021) and Samuel Bowman and George Dahl, 'What Will it Take to Fix Benchmarking in Natural Language Understanding?' in Proceedings of the 2021 Conference of the North American Chapter of the Association for Computational Linguistics: Human Language Technologies (Association for Computational Linguistics 2021) 4843

[63] Indeed, the evaluation methods previously designed for symbolic systems have not been carried over to machine learning. See further Václav Janeček and Giovanni Sartor, 'Legal Interpretation and AI: From Expert Systems to Argumentation and LLMs' (arXiv, 5 March 2026) https://arxiv.org/abs/2603.05392 accessed 15 June 2026.

[64] Jared Kaplan and others, 'Scaling Laws for Neural Language Models' (2020) arXiv:2001.08361 and Jordan Hoffmann and others, 'Training Compute-Optimal Large Language Models' (2022) arXiv:2203.15556.

[65] Zijun Yao and others, 'Are Reasoning Models More Prone to Hallucination?' (2025) arXiv:2505.23646.

Third, time has shown that what is measured through benchmarks is what gets built. Benchmarks hence do not merely record capability but rather also they steer its development.[68] This implies that if the available legal benchmarks reward classification, extraction and retrieval, model and product development is pulled towards those tasks and away from doctrinal legal reasoning. A measure that stands for the wrong target ends up entrenching the wrong target.[69] Benchmarks able to genuinely test for doctrinal legal reasoning will hence assist the development of related AI systems.[70] The absence of doctrinal-reasoning benchmarks for EU law hence also has an industrial policy dimension. Model capabilities that can be measured attract optimisation, investment and publication. Considering that existing legal benchmarks are overwhelmingly built on American materials[71] and US-law tasks, the predictable consequence is that innovation in the legal domain will be developed for, and validated against, other legal orders first. The EU's ambition to foster trustworthy AI innovation within the internal market is thus undermined by measurement infrastructure: if EU law cannot be scored innovation flows to the jurisdictions that can be. [72]

Fourth, creating a doctrinal legal reasoning benchmark will inevitably require collaboration between computer scientists and lawyers. This will, in turn, force the latter to more explicitly engage with the methods they use. The efforts needed to create a doctrinal legal reasoning benchmark for EU law will inevitably also benefit European legal methodology in its own right, whatever its eventual use in evaluating machines.

Finally, it was seen above that doctrinal reasoning is not merely one legal task among many. It rather constitutes the methodological core shared across the academy, the judiciary and practice, and a system capable of performing it well would not merely automate an isolated facet of legal work but reshape the discipline itself. The consequences would extend beyond discrete tasks to how law is created, interpreted and enforced, reaching the very structures that hold together systems governed by the rule of law.[73]

---

[66] Emily Bender and Alexander Koller, 'Climbing towards NLU: On Meaning, Form, and Understanding in the Age of Data' in Proceedings of the 58th Annual Meeting of the Association for Computational Linguistics (Association for Computational Linguistics 2020) 5185.

[67] Varun Magesh and others, 'Hallucination-Free? Assessing the Reliability of Leading AI Legal Research Tools' (2025) 22(2) Journal of Empirical Legal Studies 216.

[68] Simon Ott and others, 'Mapping Global Dynamics of Benchmark Creation and Saturation in Artificial Intelligence' (2022) 13 Nature Communications 6793, 1 ("benchmarks do not only measure, but also steer progress in AI").

[69] David Manheim and Scott Garrabrant, 'Categorizing Variants of Goodhart's Law' (2018) arXiv:1803.04585.

[70] Whether this is desirable, and what the related consequences might be, are topics for another paper. This study takes for granted that the development of both general-purpose and specialised legal AI models has already answered the question: these products will inevitably be developed.

[71] Doghe and others, above.

[72] On this goal, see further Article 1(1) AIA.

### B. The AI Act's Measurement Problem: Article 6(3), Article 15 and the Missing Benchmarks

The relationship between benchmarks and the AI Act has not yet been examined. Yet a close reading of the regulation reveals that it creates two distinct incentives to develop benchmarks that test doctrinal legal reasoning, including in EU law. First, such benchmarks are needed to test whether some AI systems can benefit from the derogation to the high-risk classification under Article 6(3) AIA. Second, for systems that cannot rely on this derogation, they are needed to implement the essential requirements for high-risk AI systems ("HRAIS") of accuracy and robustness in Article 15(1).

#### 1. The Derogation for Non-High-Risk-Systems in Article 6(3) as an Indirect Incentive for Doctrinal Legal Reasoning Benchmarks

The AI Act governs different kinds of AI systems based on their perceived levels of risk. This includes its regime for HRAIS, which applies to AI systems that meet the classification criteria outlined in Articles 6(1) and (2).[74] Article 6(2) classifies as high-risk systems that fall within the areas listed in Annex III. This includes "AI systems intended to be used by a judicial authority or on their behalf to assist a judicial authority in researching and interpreting facts and the law and in applying the law to a concrete set of facts" are HRAIS.[75] The area also catches alternative dispute resolution to the extent that the outcome produces legal effects.[76] These are, in essence, doctrinal reasoning models: systems intended to be used by a judicial authority, or on its behalf, to research and interpret the facts and the law and to apply that law to a concrete set of facts. Article 6(2) classifies AI systems falling within the areas listed in Annex III as presumptively high-risk. Article 6(3) derogates from that presumption. Such a system is not to be considered high-risk where it does not pose a significant risk of harm to the health, safety or fundamental rights of natural persons, including by not materially influencing the outcome of decision-making.

Article 6(3) qualifies as non-high-risk AI systems that do not pose a significant risk of harm to the health, safety or fundamental rights of natural persons, "including by not materially influencing the outcome of decision making" (the provision's chapeau), provided that one of four conditions is satisfied. Those conditions are that the system is intended to only (a) perform a narrow procedural task; (b) improve the result of a previously completed human activity; (c) detect decision-making patterns or deviations without replacing or influencing the previously completed assessment, subject to proper human review; or (d) perform a preparatory task to an Annex III assessment.[77] A system meeting any of these conditions nevertheless qualifies as high-risk where it profiles natural persons.[78]

[73] Contrast this with single-task benchmarks, that may touch very important domains and may equally affect individual professions while raising intricate ethical questions without, however, touching the systemic dimension a doctrinal legal reasoning benchmark would touch. Examples of the former are, inter alia, Arnaud Arindra Adiyoso Setio and others, 'Validation, Comparison, and Combination of Algorithms for Automatic Detection of Pulmonary Nodules in Computed Tomography Images: The LUNA16 Challenge' (2017) 42 Medical Image Analysis 1 and Joshua L Ebbert and Dennis Della Corte, 'PANDA-PLUS-Bench: A Clinical Benchmark for Evaluating the Robustness of AI Foundation Models in Prostate Cancer Diagnosis' (2026) 1(2) AI in Medicine 14.

[74] See further Chapter III AIA and related commentary in Michèle Finck, "The EU Artificial Intelligence Act: A Commentary (Oxford University Press 2026), Section 4.01 ff.

[75] Point 8(1) of Annex III AIA.

[76] Recital 61 AIA.

Doctrinal-reasoning benchmarks would supply evidence that an AI system falls within the scope of Article 6(3) AIA. A provider invoking the derogation must establish not only that its system fits one of the conditions in points (a)–(d) but also that, under the chapeau, it does not materially influence the outcome of decision making. According to Recital 53, this means that the system has no impact on the substance, and thereby the outcome, of decision-making, whether human or automated.[79] In the judicial context, substance is precisely what doctrinal reasoning produces. The difficulty is that classification turns on the system's intended purpose[80] as defined by the provider while the chapeau asks a question about what the system does. Benchmarks could bridge that gap. A provider who declared the purpose of the HRAIS to be, say, the retrieval and synthesis of authorities for human review could demonstrate on a doctrinal-reasoning benchmark that the system lacks the very capability whose exercise would materially influence judicial decision-making. This could inform the assessment the provider must document under Article 6(4) before placing the system on the market and into the registration that follows.[81]

Doctrinal reasoning benchmarks would also be helpful for market surveillance authorities. Article 80 empowers these authorities to evaluate a provider's Article 6(3) classification where they have sufficient reason to consider that the system was wrongly declared non-high-risk.[82] In such instances, market surveillance authorities can evaluate the AI system concerned in respect of its classification as a high-risk AI system. Testing on a benchmark would be very helpful here, especially since the powers listed in Article 74(12) are explicitly only available to market surveillance authorities dealing with high-risk AI systems. Common benchmarks used between Member States could moreover help ensure the uniform enforcement of Article 6(3).

### 2. Article 15 AIA and the Accuracy Requirement for High-Risk AI Systems

If, by contrast, an AI system used in the judicial domain meets the criteria in Point 8(a) of Annex III and the derogation in Article 6(3) is unavailable, its provider must comply with all essential requirements applicable to HRAIS.[83] This includes the accuracy, robustness, and cybersecurity requirements in Article 15(1). Benchmarks play an explicit role in this regime. Indeed, Article 15(2) directs the Commission, "in cooperation with relevant stakeholders and organisations such as metrology and benchmarking authorities," to "encourage, as appropriate, the development of benchmarks and measurement methodologies" addressing the technical aspects of how accuracy and robustness (but not cybersecurity) are to be measured. Article 15(3) additionally requires the levels of accuracy and the relevant accuracy metrics (but not robustness) to be declared in the instructions for use accompanying the system.[84]

[77] Article 6(3) AIA.
[78] Article 6(3) AIA. According to Recital 53 AIA, the notion of profiling out to be interpreted in accordance with Article 4(4) GDPR.
[79] Recital 53 AIA.
[80] See further Article 3(12) AIA.
[81] Articles 6(4) and 49(2) AIA.
[82] Article 80(1) AIA.
[83] See further Section 2 of Chapter III AIA.

Article 15 is the only provision that expressly assigns benchmarks a role in the high-risk regime. Focusing on accuracy, it requires HRAIS to achieve an appropriate level of accuracy and directs the Commission to encourage the development of benchmarks and measurement methodologies through which that requirement can be assessed.[85] Like all essential requirements, accuracy is not an absolute standard. Rather, it must be determined considering the system's intended purpose and "the generally acknowledged state of the art on AI and AI-related technologies".[86] Accuracy, in other words, is purpose-relative and this is where the measurement question returns. For systems within Point 8(a) of Annex III, the intended purpose is defined as assisting a judicial authority "in researching and interpreting facts and the law and in applying the law to a concrete set of facts". These terms map directly onto doctrinal legal reasoning. What "accuracy" means for such a system is anything but self-evident. Indeed, a transcription system's accuracy has a natural unit whereas the accuracy of legal interpretation is precisely the internal, normative, contestable and coherence-dependent quality this paper has described above. At present, no existing metric captures it. It follows that when the Commission activates Article 15(2) the benchmarks it must encourage for point 8(a) systems are doctrinal-reasoning benchmarks.

The activation of Article 15(2) by the Commission will also raise interesting questions about their status in the AIA compliance architecture. Indeed, compliance with the essential requirements runs first and foremost through harmonied standards.[87] These standards are presently being developed in response to the Commission's 2023 standardisation request.[88] Given that Article 15(2) has not yet been activated, standards are very likely to become available before any benchmarks. Once standards are available, the presumption of conformity will create an extremely strong incentive for providers to rely on standards as opposed to any other instrument. This raises the question of the practical significance of any future benchmarks, compliance with which will bear no direct presumption of conformity unless the standards will directly refer to them.

In any event, Article 15(3) mandates that the levels of accuracy and the relevant accuracy metrics of HRAIS be declared in the accompanying instructions of use. Article 13(3)(b)(ii) sharpens this obligation by requiring the instructions for use to set out "the level of accuracy, including its metrics, robustness and cybersecurity referred to in Article 15 against which the high-risk AI system has been tested and validated, and which can be expected". These obligations are unconditional and apply irrespective of whether the Commission activates Article 15(2). Indeed, the reference to "tested and validated" requires that some measurement has been performed, which in turn presupposes the existence of "relevant accuracy metrics". Yet, as of now, no agreed metric for the relevant capacity exists. This is why doctrinal-reasoning benchmarks for EU law are needed. The subsequent section will show that they are not yet available.

---

[84] The reference to "metrology and benchmarking authorities" points to an institutional infrastructure rather than to industry-led benchmark development, echoing the governace concern raised in the preceding sub-section.
[85] Articles 15(1) and 15(2) AIA.
[86] Article 8(1) AIA.
[87] Article 40 AIA confers a presumption of conformity on high-risk AI systems who conform to the harmonised standards accepted by the European Commission and published in the Official Journal of the European Union pursuant to Regulation (EU) No 1025/2012
[88] Commission Implementing Decision C(2023) 3215 final of 22 May 2023 on a standardisation request to the European Committee for Standardization and the European Committee for Electrotechnical Standardization in support of Union policy on artificial intelligence.

## V. The Doctrinal-Reasoning Benchmark Gap

Whereas the argument so far has concerned what a benchmark for doctrinal legal reasoning in EU law would have to measure, this section asks whether any existing benchmark does. It surveys the most significant benchmarks for what computer scientists refer to as "legal AI"[89] and finds that none measures doctrinal legal reasoning in EU law. The point consequential: because no existing benchmark tests this capacity, none can reveal how well a large language model performs it.

Different benchmarks were built using the case law database of the European Court of Human Rights ("ECtHR"). ECtHR-CASES (2019-21) is an outcome prediction benchmark, which asks a model to predict the correct outcome of a case given its specific factual constellation. It, for instance, asks the model to determine whether there has been a violation of an article of the European Convention on Human Rights given the facts of the case.[90] The model produces a binary prediction (violation or not) together with a multi-label prediction of which Convention articles were violated and a prediction of the case's importance score by analogy to the Court's own internal ranking. Normativity, however, relates to reasoning rather than outcome. ECtHR-PCR (2023) tests the retrieval of relevant prior case law, which constitutes pre-reasoning information work that doctrinal reasoning depends on.[91]

LexGLUE (2022) assembles seven English language legal datasets into a unified evaluation suite, covering ECtHR violation prediction; classification of US Supreme Court opinions by issue area, classification of official EU legal documents; classification of contract clauses; classification of consumer-contract terms as unfair or not; and the identification of case holdings from a five-way multiple-choice prompt.[92] Importantly, none of the seven tasks tests doctrinal legal reasoning but they rather test for classification, prediction, or recognition tasks as opposed to a defensible doctrinal justification. It engages neither the participant standpoint of internalism nor the evaluative standard of normativity.

LegalBench (2023) was developed collaboratively by computer scientists and lawyers.[93] It identifies six types of legal reasoning that LLMs can be evaluated for: issue-spotting; rule-recall;

---

[89] I embrace this terminology with hesitation as to the lawyer it denotes the lawfulness of AI rather than its use in the legal domain.

[90] It should be noted that the European Convention on Human Rights is a convention of the Council of Europe, not the European Union.

[91] T.Y.S.S. Santosh, Rashid Gustav Haddad and Matthias Grabmair, 'ECtHR-PCR: A Dataset for Precedent Understanding and Prior Case Retrieval in the European Court of Human Rights' in *Proceedings of the 2024 Joint International Conference on Computational Linguistics, Language Resources and Evaluation (LREC-COLING 2024)* (ELRA and ICCL 2024). arXiv:2404.00596.

[92] Ilias Chalkidis and others, 'LexGLUE: A Benchmark Dataset for Legal Language Understanding in English' (Proceedings of the 60th Annual Meeting of the Association for Computational Linguistics, 2022) 4310.

[93] Neel Guha and others, 'LegalBench: A Collaboratively Built Benchmark for Measuring Legal Reasoning in Large Language Models' (2023) arXiv:2308.11462.

rule-application; rule-conclusion; interpretation, and rhetorical-understanding.[94] LegalBench cannot be said to test doctrinal reasoning. Its tasks are atomic by design, and its authors explicitly note that it falls short of the coherence dimension of doctrinal legal reasoning.[95] Most tasks isolate one capability, with short fact patterns and narrow rule statements. Tasks such as clause-classification or entailment tasks, are largely paralegal preparatory work rather than legal analysis per se.[96] For more complex tasks, the benchmarks tests for a binary "yes/no" answer, which does not match the contestability feature of doctrinal legal reasoning.[97] The authors themselves acknowledge that LegalBench's tasks "focus on legal reasoning questions with objectively correct answers" and that it is thus not helpful for evaluating legal reasoning involving degrees of correctness or tasks where "reasonable minds may differ."[98] This, however, is exactly what doctrinal legal reasoning, in particular its contestability feature, requires. The benchmark is focused on American contract law.[99]

Benchmarks have, unsurprisingly, grown more sophisticated over the past two years. GreekBarBench (2025) evaluates LLMs on open-ended questions from five areas of the Greek Bar exams, scoring free-text reasoning and citation accuracy through an LLM-as-a-judge.[100] The format captures the open-ended structure of doctrinal reasoning, but because each answer is graded against a single reference answer, which leaves little room for the contestability of legal conclusions, that is to say for the possibility that two divergent answers may both be well-founded.

LEXam (2025) was built using law exam questions from the University of Zurich.[101] It goes into the direction of testing for doctrinal reasoning as it is built not just on multiple-choice but also open-ended exam questions and uses an LLM-as-judge mechanism.[102] This set-up engages the internalism and normativity elements of doctrinal legal reasoning by scoring free-text reasoning. Yet law-school examinations are pedagogical instruments, designed to test what a particular course has taught rather than to elicit a complete doctrinal analysis of a given question.[103] Two consequences follow. First, they cover a deliberately narrow range of materials: a student sitting a family-law examination is examined on the material taught in that course, not on the related property, tax, or human-rights questions that belong to other courses. LEXam therefore cannot test the coherence element of doctrinal reasoning: the requirement that a reading be made to fit the legal system as a whole. Second, because it scores answers against reference answers, it cannot test contestability, where the mark of competence is a defensible reading rather than the single right one. [104] To this may be added that the benchmark is built predominantly on Swiss law, so that it bears on EU law only indirectly.

[94] Ibid, page 6.
[95] LegalBench, page 10 ("Finally, LEGALBENCH evaluates IRAC abilities independently, while law exams and other legal work requires lawyers to generate outputs which follow IRAC in a multi-hop matter (i.e., each aspect is applied to the same fact pattern")
[96] See further Neel Guha and others, 'Contract NLI: Notice on Compelled Disclosure' (LegalBench, 2023) https://hazyresearch.stanford.edu/legalbench/tasks/contract_nli_notice_on_compelled_disclosure.html accessed 16 June 2026.
[97] Ibid and Neel Guha and others, 'Hearsay' (LegalBench, 2023) https://hazyresearch.stanford.edu/legalbench/tasks/hearsay.html accessed 16 June 2026.
[98] Ibid, page 10.
[99] Ibid.
[100] Odysseas S Chlapanis and others, 'GreekBarBench: A Challenging Benchmark for Free-Text Legal Reasoning and Citations' (2025) arXiv:2505.17267.
[101] Yu Fan and others, 'LEXam: Benchmarking Legal Reasoning on 340 Law Exams' (2025) arXiv:2505.12864, 2.
[102] Ibid.
[103] Indeed the prompt for the open exam questions explicitly required the LLM to answer in an "structured, exam-style manner" rather than the more comprehensive doctrinal reasoning style that may be needed regarding the particular question at issue. See further ibid, 29. Similarly, the system prompt for the LLM-as-judge also required it to act as a "judge specializing in the evaluation of Swiss law school exams". See further ibid, 30.

BenGER (2026) is a benchmark for subsumption-based legal reasoning in German law. The benchmarks uses the internal method and is indeed specifically built to catch a model that produces the appearance of internal reasoning without performing it. BenGER indeed "valuation as a distribution over plausible expert assessments rather than a single deterministic label".[105] This acknowledges the normative and contestable nature of doctrinal legal reasoning. The benchmark however does not meet the coherence part of what doctrinal legal reasoning would account for when testing knowledge over an entire legal system, such as EU law. It tests the structured application of a rule to a set of facts, not whether a reading coheres with the rest of the legal system. The benchmark is built on the German four-step case-solving scaffold: state the applicable norm, define its elements, apply them to the facts, and conclude.[106] What it scores is how well each element of the chosen norm is tied to the facts[107], marking each step of the scaffold as a separate component "rather than to a global quality impression".[108] It follows that BenGER does not measure coherence in the doctrinal sense: whether the reading is consistent with, and supported by, the wider body of law.

Around the time this paper was first published as a pre-print, the LegalTech start-up Harvey released its Harvey LAB benchmark.[109] This benchmark, however, merely measures how models perform on "real-world legal tasks". The purpose of the benchmark is to measure how much of a lawyer's task a model can finish, not whether the reasoning behind the work is sound.[110] That measures useful output, not doctrinal defensibility.

This section established the absence of benchmarks for doctrinal legal reasoning in EU law. The subsequent section offers a corresponding constructive contribution. It presents a taxonomy of failure modes designed to define what a benchmark must detect to genuinely evaluate EU doctrinal reasoning.

## VI. A Failure Taxonomy for EU Doctrinal Reasoning

Early on, the literature on AI for the legal domain was largely focused on hallucinations such as fabricated cases, invented citations, propositions with no existence in any source.[111] Initially, such hallucinations were so obvious that no one was asking whether LLMs can genuinely reason doctrinally. Indeed, hallucinations are a machine-specific pathology that does not have systemic analogous shortcomings in human legal work. It is, moreover, a shortcoming detectable by simple lookup, which is why retrieval-augmented architectures have significantly (though not entirely) reduced hallucination rates.[112]

---

[104] Page 2.
[105] Sebastian Nagl and others, 'BenGER: Benchmarking LLM Systems on Subsumption-Based Legal Reasoning in German Law' (2026) arXiv:2605.28183, 1.
[106] Ibid, 1.
[107] Ibid, 2 ("the rubric rewards how each norm element is anchored in the specific fact pattern").
[108] Ibid.
[109] Niko Grupen, Gabe Pereyra and Julio Pereyra, 'Introducing Harvey's Legal Agent Benchmark' (Harvey, 6 May 2026) https://www.harvey.ai/blog/introducing-harveys-legal-agent-benchmark accessed 16 June 2026.
[110] Ibid.
[111] Dahl et al (n xx).

The failures outlined below are of a different nature. Each is a mistake that competent human lawyers are also capable of making: they mark the boundary between doctrinal competence and its absence, the errors that legal education and professional apprenticeship exist to train out. These failures are, moreover, not detectable by mere lookup. Rather, detecting them requires doctrinal competence itself. A system robust against this taxonomy is not simulating doctrinal reasoning but performing something measurably equivalent to it.

Two preliminary points frame the taxonomy. The first concerns what it is that we should be evaluating. Currently, AI for use in the legal domain generally rely on retrieval-augmented generation.[113] This means that rather than answering from the model's training alone, the system first searches a body of legal sources and passes what it retrieves to the model, which then composes the answer. The proper object of evaluation is therefore the model together with the retrieval layer, not the model in isolation. That distinction matters because the retrieval step builds in specifically legal distortions that the model's later reasoning cannot detect or repair. To be made searchable, source documents are first broken into short fragments, which detaches an individual provision from the instrument and the wider framework in which it sits, stripping away the systematic context doctrinalism depends on.[114] The retrieval mechanism then ranks those fragments by topical resemblance rather than legal authority, surfacing passages that merely look similar in wording or subject matter instead of those that are binding or genuinely on point. And because the system must convert text into numerical representations to compare it, the multilingual character of EU law is flattened, thus reducing cross-lingual legal nuance before any doctrinal analysis can engage with it.[115]

The taxonomy comprises twenty-one failure. It is the contribution a legal scholar can responsibly make to a project that will ultimately require interdisciplinary collaboration. It is neither exhaustive nor a finished benchmark design. Each failure is one a competent EU lawyer would recognise as disqualifying, and that no current benchmark detects, and that bears directly on the "appropriate" accuracy required of high-risk AI systems deployed in the judicial domain.

---

[112] Qianchi Zhang and others, 'Stable-RAG: Mitigating Retrieval-Permutation-Induced Hallucinations in Retrieval-Augmented Generation' (arXiv, 2026) https://arxiv.org/abs/2601.02993 accessed 15 June 2026; Cheng Niu and others ‘RAGTruth: A Hallucination Corpus for Developing Trustworthy Retrieval-Augmented Language Models’ in Proceedings of the 62nd Annual Meeting of the Association for Computational Linguistics (2024) 10862–10878, Association for Computational Linguistics; Yujia Zhou and others, 'Trustworthiness in Retrieval-Augmented Generation Systems: A Survey' (arXiv, 2024) https://arxiv.org/abs/2409.10102 accessed 15 June 2026.

[113] Magesh and others (n xx).

[114] Markus Reuter and others, 'Towards Reliable Retrieval in RAG Systems for Large Legal Datasets' in Nikolaos Aletras and others (eds), Proceedings of the Natural Legal Language Processing Workshop 2025 (Association for Computational Linguistics 2025) and Andrea Ferraris and others, 'Legal Chunking: Evaluating Methods for Effective Legal Text Retrieval' in Legal Knowledge and Information Systems: JURIX (Frontiers in Artificial Intelligence and Applications 2024)

[115] Ilias Chalkidis, Manos Fergadiotis and Ion Androutsopoulos, 'MultiEURLEX — A Multi-lingual and Multi-label Legal Document Classification Dataset for Zero-shot Cross-lingual Transfer' in Proceedings of the 2021 Conference on Empirical Methods in Natural Language Processing.

## A. Failures of Source Recognition and Authority

The failures grouped below concern an operation on which all doctrinal reasoning depends, namely recognising the source of an instrument and the authority attached to it. Section III exposed that doctrinal reasoning is internal, and internalism entails situating the norm in the system's own hierarchy.

1. *Errors related to the hierarchy of norms*. Treating norms of different rank as equivalent or inverting their rank is a mistake that disqualifies subsequent reasoning. For instance, assigning a soft law instrument the value of a Treaty provision or overlooking that the Charter has the same legal value as the Treaties while a directive does not will discredit the human reasoner and its output.[116]

2. *Confusing different layers of norms*. Competent doctrinal reasoning must also situate that norm within the layered hierarchy of norms that constitutes EU law. For instance, *Mangold* established that the right to non-discrimination on grounds of age exists as a general principle of EU law and that this right has horizontal direct effect between two private parties.[117] The right also has the status of primary law through Article 21(1) CFR. In addition, it finds expression in secondary legislation, specifically the Equal Treatment Directive.[118] Similarly, the right to data protection is anchored in primary law in Articles 8 CFR and 16 TFEU and then concretised by secondary legislation in the form of the General Data Protection Regulation.[119] Doctrinal legal reasoning must not just engage with the substantive right but also the layer in which it is anchored.[120] Indeed, the norm's status in this layered architecture will affect crucial questions such as whether it is capable of horizontal direct effect, what its specific scope is; its validity as primary-law norm can be used to test the validity of secondary law, not the reverse; the impact of consistent interpretation or its temporal reach.[121]

3. *Treating soft law as a unitary whole*. Soft law matters in EU law, yet its effect diverges widely depending on the specific are of law in question. Soft law has been defined as "rules of conduct that are laid down in instruments which have not been attributed legally binding force as such, but nevertheless may have certain (indirect) legal effects, and that are aimed at and may produce practical effects".[122] These effects, however, vary significantly depending on the specific area. In competition law, soft law carries significant weight, influenced by it being an exclusive EU competence where the Commission acts as the enforcer.[123] In data protection law the guidelines of the European Data Protection Board and, formerly, the Article 29 Working Party, carry significant weight in data protection practice, yet "the Court never cites WP29 opinions in its data protection jurisprudence, and on a number of occasions has ruled contrary to the WP29's earlier opinions".[124] The influence of soft law hence derives from procedural and institutional circumstances rather than from anything explicitly encoded in the text, which means that models must determine the jurisprudential weight of each instrument for themselves, just as human doctrinal reasoning does.

---

[116] Article 6(1) TEU. See further Deirdre Curtin and Tatevik Manucharyan, 'Legal Acts and Hierarchy of Norms in EU Law', in Anthony Arnull and Damian Chalmers (eds), The Oxford Handbook of European Union Law (OUP 2015), 103-125 and Jacques Ziller, 'Hierarchy of Norms: Hierarchy of Sources and General Principles in European Union Law' in Ulrich Becker and others (eds), *Verfassung und Verwaltung in Europa: Festschrift für Jürgen Schwarze zum 70. Geburtstag* (Nomos 2014) 334.

[117] Case C-144/04 *Mangold* [2005] ECLI:EU:C:2005:709, para 75. See also Case C-555/07 *Seda Kücükdeveci* [2010] ECLI:EU:C:2010:21 and Mirjam de Mol, 'Kücükdeveci: Mangold Revisited – Horizontal Direct Effect of a General Principle of EU Law' (2010) 6 European Constitutional Law Review 293.

[118] Article 1 of Council Directive 2000/78/EC of 27 November 2000 establishing a general framework for equal treatment in employment and occupation [2000] OJ L303/16.

[119] Regulation (EU) 2016/679 of the European Parliament and of the Council of 27 April 2016 on the protection of natural persons with regard to the processing of personal data and on the free movement of such data, and repealing Directive 95/46/EC (General Data Protection Regulation) [2016] OJ L119/1.

[120] Regulation (EU) 2016/679 of the European Parliament and of the Council of 27 April 2016 on the protection of natural persons with regard to the processing of personal data and on the free movement of such data, and repealing Directive 95/46/EC (General Data Protection Regulation) [2016] OJ L119/1 (hereafter "GDPR").

[121] See further Article 51(1) CFR.

4. *Misweighting judicial authority*. The output of the CJEU can only be fully understood if it is contextualised in judicial hierarchy. This implies that failing to register the differential weight of pronouncements within the EU judicial hierarchy, including the distinction between Grand Chamber and chamber judgments, the appeal status of General Court rulings, and the non-binding character of Advocate General opinions leads to doctrinal reasoning that is flawed.[125] LLMs that only account for the statistically most likely text associations will miss that an AG Opinion may have been extensively quoted before the related judgment, yet which was not followed by the subsequent Court decision.

5. *Disregarding the autonomous nature of EU law concepts*. It was seen above that concepts of EU law must be given an autonomous meaning. It follows that they do not necessarily carry the same meaning as their counterparts in the Member States' legal systems, still less in other legal systems. A model whose internal representations cannot distinguish between these meanings will produce incorrect answers to legal questions

**B. Failures of EU Law's Operative Doctrines**

The failures outlined in this section concern the doctrines that determine whether and how an EU norm produces legal effects, against whom it may be invoked, and with what authority it is endowed. Doctrines such as supremacy, direct effect, direct applicability, consistent interpretation, and the differentiated effects attached to the Court's jurisdiction are an expression internalism. They form part of what a competent doctrinal reasoner knows, yet are not explicitly expressed in legal text. This implies that a model ingesting substantive norms as a flat corpus misses the structure that is key to understanding and correctly applying these norms.

6. *Errors related to the effects of a doctrine*. These errors entail missing or confusing the doctrines of direct effect, direct applicability, consistent interpretation and state liability, or misstating against whom a provision can be invoked. An of this error class would be misjudging when a provision of EU law is capable of horizontal direct effect – one of the classical exam questions used in EU law classes to test whether students grasp the nuanced doctrines of EU law.[126] Just like these students, a competent model would need to account for the type of norm, the identity of the parties (are they the state, an emanation of the state or a private party?) to correctly assess which doctrines apply in a given setting.

---

[122] Linda Senden, 'Soft Law and its Implications for Institutional Balance in the EC' (2005) 1 Utrecht Law Review 79, at 81. See also Article 288 TFEU and Case C-322/88 *Grimaldi* [1989] ECR 4407 as well as Emilia Korkea-aho, 'National Courts and European Soft Law: Is *Grimaldi* Still Good Law?' (2018) 37 Yearbook of European Law 470.
[123] Zlatina Georgieva, 'Soft Law in EU Competition Law and its Judicial Reception in Member States: A Theoretical Perspective' (2015) 16 German Law Journal 223.
[124] Nadezhda Purtova "The Law of Everything. Broad Concept of Personal Data and Future of EU Data Protection Law" (2018) Law, Innovation and Technology 40.
[125] Article 252 and 256 TFEU and Articles 16 and 56-61 of the Statute of the CJEU.

7. *Procedural errors.* The competent doctrinal reasoner knows that the case law of the CJEU cannot be treated as a single, fungible, corpus but that rather the weight, reach and effect of a judgment depends on the procedure under which it was rendered. As such, a preliminary ruling issued under the related procedure foreseen in Article 267 TFEU binds the referring court and form an authoritative interpretation of EU law that generally applies *erga omnes*.[127] This is distinct from an infringement judgment under Article 258 TFEU, which is declaratory and does not automatically create individual rights.[128] Doctrinally competent LLMs need to differentiate the procedural contextuality of these judicial outputs, which is not made explicit in its text.

8. *Errors related to the scope of application of EU law*. EU law norms differ in their legal effect due to their scope. This implies that the competent doctrinal reasoner must be able to assess when a situation is governed by EU law, such as in, for instance, determining when Charter rights apply[129] or assessing when there is a purely internal situation.[130] Competent doctrinal reasoning models are hence those that do not generalise topically and produce Charter analysis whenever "fundamental rights" are mentioned. At the same time, an answer that treats a legal question as one involving only Member State law when EU law in fact applies would be equally disqualifying.

9. *Blindness towards national implementation*. EU law knows two distinct sources of secondary legislation: regulations and directives.[131] Regulations have direct application, are binding in their entirely and directly applicable in all Member States.[132] In contrast, directives are merely binding "as to the result to be achieved, upon each Member State to which it is addressed, but shall leave to the national authorities the choice of form and methods".[133] This implies that Member States must implement directives and that different Member States do so in different manners. As a result, there is less normative uniformity between Member States and the national implementation adds nuance to the terms of the directive itself. A human reasoner taking the text of the directive to assess national effects *inter partes* would disqualify herself. Models equally need to acknowledge that the answer turns on national transposition.

---

[126] See further Case 152/84 Marshall EU:C:1986:84; Case C-91/92 Faccini Dori EU:C:1994:292. Case C-414/16 Egenberger EU:C:2018:257; Joined Cases C-569/16 and C-570/16 Bauer EU:C:2018:871; Case C-684/16 Max-Planck EU:C:2018:874

[127] Case 66/80 *International Chemical Corporation* [1981] ECR 1191. See further Giuseppe Martinico, 'Retracing Old (Scholarly) Paths. The Erga Omnes Effects of the Interpretative Preliminary Rulings' (2023) 15 European Journal of Legal Studies 37.

[128] See also Article 260(1) TFEU.

[129] Article 51(1) CFR and Case C-617/10 *Åklagaren v Hans Åkerberg Fransson* EU:C:2013:105.

[130] Case C-268/15 *Ullens de Schooten* EU:C:2016:874 and Sara Iglesias Sánchez, 'Purely Internal Situations and the Limits of EU Law: A Consolidated Case Law or a Notion to be Abandoned?' (2018) 14 European Constitutional Law Review 7.

[131] Article 288 TFEU.

[132] Ibid.

[133] Ibid.

## C. Failures of Interpretive Method

The failures grouped here concern interpretative method, how a reading of the law is reached, and what makes one reading correct. Correctly interpreting EU law is not simply a matter of reading the text faithfully and reflecting what the sources say so that a model which had absorbed the whole body of EU legal materials would, as a matter of course, interpret it correctly. Indeed, EU doctrinal reasoning is normative. It is settled by the better argument towards the order's purposes and by the normative weight each source carries, neither of which is made explicit in the text or depends on how often a reading recurs. A model that merely reproduces the reading most frequent in its training material will hence fail at doctrinal competence.

10. *Failure of method selection*. Whereas EU legal interpretation is famously teleological, teleology is not uniformly required.[134] Rather, the competent interpreter judges when literal meaning suffices, when context matters, and when the supranational legal order's objectives must be accounted for. If the latter scenario arises, she must identify the purposes of EU law that matter for legal interpretation. ´

11. *Mistaking frequency for correctness*. EU law has changed over time, as any legal order does. For instance, in *Dassonville* the Court read the prohibition on measures having equivalent effect to quantitative restrictions (now Article 34 TFEU) expansively, catching any national rule capable of hindering intra-Union trade directly or indirectly, before drawing back in *Keck*, which placed "certain selling arrangements" outside Article 34 altogether where they apply to all traders and affect domestic and imported goods alike.[135] A model trained mostly on pre-*Keck* commentary, or on a corpus dominated by sources that simply restate *Dassonville*, will reproduce the pre-*Keck* position and do so confidently but incorrectly.

12. *Misunderstanding recitals*. Understanding the legal status of recitals in EU law requires nuanced reflection. Indeed, recitals have, formally speaking, no independent legal status but are merely intended to guide the interpretation of operative provisions of secondary legislation. Yet, there is a growing trend of preambles to secondary legislation becoming increasingly expansive and the Court has, arguably, relied on them beyond what is formally foreseen in basing specific rights on a recital.[136] LLMs hence need to be able to distinguish recitals and an operative article, which sit in the same document and read alike while nothing on the face of the text marks the one as explanatory and the other as binding. Furthermore, they also need to understand in which cases and areas of law the Court has, in fact, given more interpretative weight than it formally should.[137]

13. *Failure to consider multilingualism*. It was seen above that EU law is equally authentic in twenty-four language versions and no single version may serve as the sole basis of interpretation or be given automatic precedence. Models treating the English text of an EU instrument as the interpretive baseline will fail at doctrinal excellence in the cases where different language versions diverge.[138]

---

[134] See further Section III above.

[135] Joined Cases C-267/91 and C-268/91 *Keck and Mithouard* EU:C:1993:905.

[136] Case C-136/04 *Deutsches Milch-Kontor GmbH v Hauptzollamt Hamburg-Jonas* [2005] and Article 296(2) TFEU; Joint Practical Guide of the European Parliament, the Council and the Commission for Persons Involved in the Drafting of European Union Legislation (2nd edn, Publications Office of the European Union 2015). The legal status of recitals will be particularly challenging in respect of the AI Act considering the frequent discrepancies between recitals and operative provisions in this instrument. On this, see further further Michèle Finck, „The EU Artificial Intelligence Act: A Commentary" (OUP 2026) at 22.

[137] Case C-634/21 *OQ v Land Hessen* (*SCHUFA Holding*) EU:C:2023:957, paras 45 and 53.

### D. Failures of Temporal and Contested Reasoning

The failures grouped in this section concern the law as a moving, contested target. A doctrinal practitioner reasons against interpretation that evolves, regimes that update, and questions that are the subject of disagreement. Likewise, a doctrinally competent model cannot answer with equal confidence on a point that is settled, one which is not and one that the law has moved past.

14. *Misstatements of jurisprudential evolution*. The competent doctrinal reasoner cannot present a position as current when it has been narrowed, refined or overruled, such as in assuming that *Dassonville* has not been affected by *Keck*, contrary to what was discussed just above.

15. *Misunderstanding the temporal evolution of legal landscapes*. The competent doctrinal reasoner must account for the way the law evolves over time. Article 15(1) AIA, examined above, illustrates the point. At present, what its accuracy requirement demands is settled by reading the provision itself in its legislative setting, including Article 8(1). Once harmonised standards, and, potentially, the benchmarks become available, the same exercise must be informed by those instruments. In due course, case law also needs to be accounted for. The provision itself will not have changed, but the body of material against which its meaning is fixed will have grown, so that a reading which was adequate at one moment becomes incomplete at the next.

16. *Failure to flag contestation.* The competent doctrinal reasoner is aware of ongoing disputes in EU law, such as pending references for preliminary ruling, scholarly disagreements and bumpy judicial dialogues. Even with up-to-date retrieval a model may fail to recognise that a pending reference could unsettle what would otherwise be *acte clair*. Pursuant to this doctrine, a national court of last instance, otherwise obliged by Article 267(3) TFEU to refer a question of EU law for preliminary ruling to the CJEU, is exceptionally relieved of that obligation where the correct application of EU law is so obvious as to leave no scope for any reasonable doubt, provided the court is convinced the matter would be equally obvious to the courts of the other Member States and to the Court itself.[139]

17. *Failures to locate boundaries*. It has been seen above that contestability is one of the four core features of doctrinal legal reasoning. The doctrinal legal reasoner has failed where she produces a single confident answer to a question that is *acte clair*. LLMs, known for being overly confident on matters where it is not warranted, must equally master such nuance.[140]

---

[138] Case 29/69 *Erich Stauder* [1969] ECR 419, para 3.
[139] Case 283/81 *CILFIT* [1982] ECR 3415, paras 16–21.
[140] Katherine Tian and others, 'Just Ask for Calibration: Strategies for Eliciting Calibrated Confidence Scores from Language Models Fine-Tuned with Human Feedback' (2023) Proceedings of the 2023 Conference on Empirical Methods in Natural Language Processing 5433, arXiv:2305.14975 and Miao Xiong and others, 'Can LLMs Express Their Uncertainty? An Empirical Evaluation of Confidence Elicitation in LLMs' (2024) Proceedings of the Twelfth International Conference on Learning Representations, arXiv:2306.13063.

## E. Failures of Coherence

The failures grouped here concern coherence, the demand that a doctrinal account hang together as a whole and that its conclusions rest on the reasons given for them, in line with what was observed above. Indeed, coherence is not mere consistency, the absence of contradiction. It rather is the positive property of propositions hanging together into a systematically ordered account, and it is what decides hard cases, where two readings are each consistent with the authoritative materials and the interpreter prefers the one that better fits the principles underlying the body of law.

18. *Failure to integrate systemically.* A doctrinal reasoner should not produce locally correct but globally incorrect analysis. Indeed, the coherence component of doctrinarism that was introduced above requires that between two alterative readings, the one coherent with the legal order overall be chosen. This is a challenge for LLMs as the "lost in the middle" literature shows that model performance on tasks requiring integration of information across long contexts degrades substantially when relevant information is not at the beginning or end of the context window.[141]

*29. Incoherence across instruments.* Doctrinal legal reasoning must account for all the instruments applicable to a scenario, not only the one under which the question is posed. This is particularly pressing in digital regulation, as the digital acquis is composed of many distinct instruments, each complex and each overlapping with its neighbours in ways that are sometimes contradictory. Yet a doctrinally competent answer to a question under one instrument cannot contradict the instruments adjacent to it.[142] Cross-instrument consistency is, however, not tracked by retrieval-augmented systems, which return the most relevant chunk for each sub-question without checking that the chunks' implications are mutually consistent.[143]

20. *Inflating authority*. This error arises when non-binding materials are treated as binding, such as assuming that an Advocate General opinion, even where not followed by the Court is considered as binding. The distinction is especially demanding in EU law, where recitals inform interpretation without carrying independent binding force and soft-law instruments such as recommendations must be taken into consideration without being binding, so that a model unable to grade these degrees of authority will present as settled law what is in truth only persuasive.

21. *Mismatching citation and proposition* One of the more subtle mistakes in human doctrinal reasoning is a real citation offered for a proposition it does not in fact support. This is also the failure mode into which retrieval-augmentation converts fabrication, and it is more dangerous than hallucinations: the citation is real, the source is real, and only a careful reading of the cited document reveals the mismatch.[144]

[141] Taiming Lu and others, 'Insights into LLM Long-Context Failures: When Transformers Know but Don't Tell' (2024) arXiv:2406.14673.

[142] See, e.g. the relationshp between Articles 22 GPDR and 86 AI Act.

[143] Varun Magesh and others, 'Hallucination-Free? Assessing the Reliability of Leading AI Legal Research Tools' (2025) 22 Journal of Empirical Legal Studies 216

[144] Jonas Wallat and others, 'Correctness is not Faithfulness in RAG Attributions' (2024) arXiv:2412.18004.

## VII. Closing the Measurement Gap

The measurement gap ultimately raises the familiar question of what can be measured and what matters at the doctrinal core of EU law. The AI Act requires high-risk judicial systems to be accurate, but what counts as accuracy for such systems cannot be defined without exactly the instrument we lack. This means that closing the gap is no longer a scholarly ambition but a legal necessity. The current gap is understandable as benchmarks are traditionally built by computer scientists, good ones are costly, and there is no settled account of what doctrinal legal reasoning requires, which may make it hard to measure at all. The paper has proceeded regardless, both because benchmarks are the tool by which machine learning currently measures progress and because the AI Act now compels their creation. This means that building one is therefore necessary, even if it may ultimately prove impossible. The contribution a doctrinal lawyer can responsibly make to that interdisciplinary task is the one that was offered in this paper: a taxonomy of the failure modes any genuine doctrinal-reasoning benchmark for EU law would have to detect.